# Frequency-dependent rigidity in large-scale interferometric gravitational-wave detectors

F.Ya.Khalili

*Dept. of Physics, Moscow State University,*
*Moscow 119899, Russia*
*e-mail: farid@mol.phys.msu.su*

## Abstract

Electromagnetic rigidity which exists in large-scale optical resonators if pumping frequency is detuned from the eigenfrequency of resonator have sophisticated spectral dependence which allows to obtain sensitivity better than the Standard Quantum Limits both for the free test mass and the harmonic oscillator.

## 1 Introduction

The standard quantum limit (SQL) [1] is one of the most fundamental factors which prevent the gravitational-wave antennae [2] sensitivity from increasing. The basis for this limit is the uncertainty relation for two kind of noises inherent in position meters: the measurement noise and the back action noise. In the interferometric position meters, these noises are proportional to phase fluctuations of the pumping beam and radiation pressure noise, correspondingly.

Several methods to overcome the SQL has been proposed (see, for example, articles [3, 4, 5, 6, 7]) but most of them encounters serious technological limitation and/or has some other disadvantages which do not permit implementation of these methods in the near future. On the other hand, the SQL itself is not an absolute limit but, in particular, it depends on the dynamic properties of the test object which is used in the experiment. The well known example is the harmonic oscillator. Its response to a resonant force is relatively strong and it allows to use less sensitive meter (with larger measurement noise and therefore with smaller back action noise). Due to this property the harmonic oscillator allows to obtain sensitivity better than the SQL for the free mass when the signal frequency is close to the eigen frequency of the oscillator $\Omega_m$ [8, 9, 10, 11].

In the articles [8, 10] it was shown that it is possible to create very low noise mechanical rigidity using Fabry-Perot resonators with detuned pumping. In the article [11] it was also shown that such a rigidity exists in the signal-recycled topology of the gravitational-wave antennae and it permits to overcome the SQL for a free mass in narrow band.

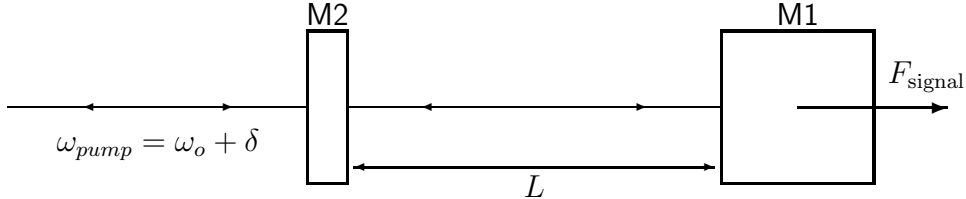

Figure 1: Simplified scheme of the interferometric detector

One can imagine the following frequency dependent rigidity:

$$K(\Omega) = m\Omega^2 \qquad (1)$$

where $m$ is the mass it is attached to, and $\Omega$ is an *arbitrary* observation frequency. In principle, such a rigidity would allow to obtain arbitrarily high sensitivity throughout the spectral range where the formula (1) is valid. On the other hand, it is well known that the mechanical rigidity created by parametrical opto-mechanical systems can be frequency dependent. In this article we show that the mechanical rigidity frequency dependence in large-scale interferometric meters with bandwidth comparable to or smaller than the signal frequency can be close to the formula (1) in some spectral band.

## 2 The simple example: Second-order-pole regime

Consider the simplified interferometric detector scheme presented in Fig. 1. Here the signal force with the amplitude $F_{\text{signal}}$ which had to be detected, acts on the test mass $m$. This mass serves also as mirror M1 and together with second mirror M2 forms Fabry-Perot resonator. In this article we suppose that refraction of the mirror M1 is equal to unity and there is no absorption in the mirror M2.

The resonator is pumped at the frequency $\omega_{\text{pump}}$ which is detuned far from its eigen frequency $\omega_o$:

$$\delta = \omega_{\text{pump}} - \omega_o \gg \gamma, \qquad (2)$$

where $\gamma$ is the half-bandwidth of the resonator. This detuned pumping creates a ponderomotive rigidity. One of the reflected beam quadrature amplitudes is measured, giving information about the mirrors relative position.

We will refer to this simple scheme in this article but it can be shown that all results obtained here are valid for the signal-recycled topology [12] planned for the second stage of the LIGO program.

By solving this system equations of motion it is easy to show that if $\gamma \to 0$ then mechanical rigidity created by the optical pumping will be equal to

$$K(\Omega) \approx \frac{2\omega_o \mathcal{E} \delta}{L^2(\delta^2 - \Omega^2)} \tag{3}$$

where $\Omega$ is the observation frequency, $\mathcal{E}$ is the optical energy stored in the resonator, and $L$ is the resonator length.

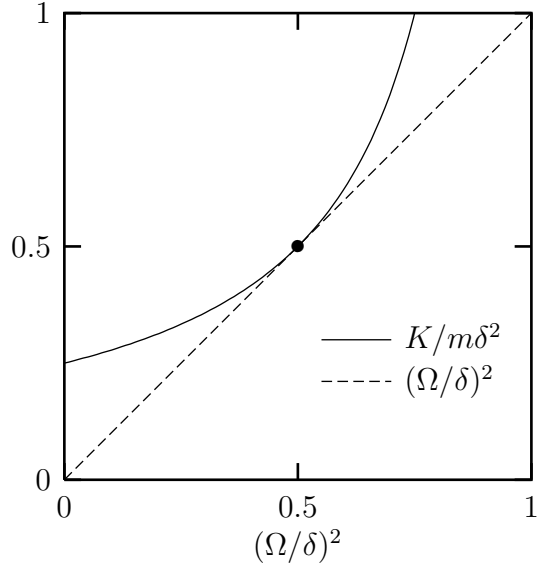

Figure 2: Second-order pole ($\gamma \to 0$)

Our goal is to set $K(\Omega)$ as close to the ideal frequency dependence (1) as possible nearby some given value of $\Omega$, so we require that

$$K(\Omega) = m\Omega^2 \tag{4}$$

and

$$\frac{dK(\Omega)}{d\Omega} = \frac{d(m\Omega^2)}{d\Omega} . \tag{5}$$

It is easy to show that these conditions can be fulfilled if (and only if)

$$\Omega = \Omega_2 \equiv \frac{\delta}{\sqrt{2}} \tag{6}$$

and

$$\mathcal{E} = \frac{mL^2\delta^3}{8\omega_o} = \frac{mL^2\Omega_2^3}{2\sqrt{2}\omega_o}. \tag{7}$$

In this case susceptibility of test object which consists of mass $m$ and such a rigidity

$$\chi(\Omega) = \frac{1}{-m\Omega^2 + K(\Omega)} \tag{8}$$

will have a second-order pole at the frequency $\Omega_2$, *i.e* if $|\Omega - \Omega_2| \ll \Omega$ then we will obtain

$$-m\Omega^2 + K(\Omega) \approx 4m(\Omega - \Omega_2)^2 \tag{9}$$

where subscript "2" means "second-order pole" (see Fig. 2, where second-order-pole point marked by "•"). It means if the signal force has the form of a sinusoidal train with duration $\tau_F$ and the mean frequency $\Omega_F \simeq \Omega_2$ then the amplitude of the mass $m$ oscillations caused by this force will be proportional to[1]

$$x_{\text{signal}} \sim \frac{F_{\text{signal}}\tau_F^2}{m} \tag{10}$$

(in this section we omit all numerical factors of the order of unity). On the other hand, it can be shown that the Standard Quantum Limit for such a second-order-pole system is equal to

$$x_{\text{SQL}}^{(2)} \sim \sqrt{\frac{\hbar\tau_F}{m}}. \tag{11}$$

Hence using this system and ordinary position meter, it is possible to detect the force

$$F_{\text{signal}} \sim \frac{mx_{\text{SQL}}^{(2)}}{\tau_F^2} = F_{\text{SQL}}^{(2)} = \sqrt{\frac{\hbar m}{\tau_F^3}}. \tag{12}$$

This value is $\Omega_F\tau_F$ times smaller than the SQL value corresponding to the free test mass

$$F_{\text{SQL}}^{\text{free mass}} = \sqrt{\frac{\hbar m \Omega_F^2}{\tau_F}} \tag{13}$$

---

[1] We want to remind that the free mass has second-order pole at zero frequency, and the harmonic oscillator has first-order pole at resonance frequency.

and $\sqrt{\Omega_F\tau_F}$ times smaller than the SQL for the harmonic oscillator with ordinary frequency-independent rigidity,

$$F_{\rm SQL}^{\rm oscillator} = \frac{\sqrt{\hbar m\Omega_F}}{\tau_F}. \tag{14}$$

It had to be noted that the energy (7) is close to the energy

$$\mathcal{E} = \frac{mL^2\Omega^3}{2\omega_o}. \tag{15}$$

which is necessary to achieve the SQL using traditional scheme of the interferometric position meter.

## 3 Sensitivity for different regimes of the frequency-dependent rigidity

In this section we will use spectral approach based on the total net noise of the meter (see article [13]). This noise is normalized in such a way that the signal-to-noise ratio is equal to

$$\frac{s}{n} = \int_{-\infty}^{\infty} \frac{|F_{\rm signal}(\Omega)|^2}{S_{\rm total}(\Omega)} \frac{d\Omega}{2\pi}, \tag{16}$$

where $F_{\rm signal}(\Omega)$ is the spectrum of the signal force, and $S_{\rm total}(\Omega)$ is the spectral density of this noise.

In the case of the interferometric detector (see Fig. 1) spectral density of the net noise is equal to

$$S_{\rm total}(\Omega) = S_F^{\rm eff}(\Omega) + \chi_{\rm eff}^{-2}(\Omega)S_x(\Omega), \tag{17}$$

where

$$S_F^{\rm eff}(\Omega) = \frac{\hbar^2}{4S_x(\Omega)} \tag{18}$$

is the residual back-action noise of the meter (*i.e.* part of the back-action noise $S_F$ which does not correlate with the measurement noise),

$$S_x(\Omega) = \frac{\hbar L^2}{8\omega_o\mathcal{E}\gamma} \times \frac{\Omega^4 + 2\Omega^2(\gamma^2-\delta^2) + (\gamma^2+\delta^2)^2}{\Omega^2+\gamma^2} \tag{19}$$

is the measurement noise,

$$\chi_{\text{eff}}(\Omega) = \frac{1}{-m\Omega^2 + K_{\text{eff}}(\Omega)} \tag{20}$$

is the effective susceptibility of the system, and

$$K_{\text{eff}} = \frac{2\omega_o \mathcal{E}\delta}{L^2} \times \frac{3\gamma^2 + \delta^2 - \Omega^2}{\Omega^4 + 2\Omega^2(\gamma^2 - \delta^2) + (\gamma^2 + \delta^2)^2} \tag{21}$$

is the effective rigidity which is the sum of two terms: (i) the real physical rigidity which exists in the system due to the dependence of the optical energy in the resonator on the mirrors position, and (ii) the "virtual" rigidity introduced by the cross-correlation of the measurement noise and back-action noise (see article [14]). It should be noted that in our case the real rigidity is much larger than the "virtual" one. However, it is the "virtual" rigidity that compensates the imaginary part of the real physical rigidity which describes dynamical instability of the system. In other words, the instability does exist and must be compensated by some feed-back scheme but the meter does not "see" it.

Expressions (19,21) are obtained for the case where the phase quadrature amplitude of the output optical wave is measured. Our results does not depend essentially on which quadrature amplitude is measured but choosing of the phase quadrature amplitude provides slightly better results and also allows to simplify the formulae.

The behavior of the $K_{\text{eff}}(\Omega)$ is rather sophisticated and allows a several different regimes, depending on the pumping energy, resonator bandwidth and detuning: with three first-order poles [Fig. 3(a)]; with one second-order and one first-order poles [Fig. 3(b,c)]; and with one third-order pole [Fig. 3(d)]. Which one should be chosen depends on the signal form. Detailed analysis of all of them exceeds the frames of this short article. One of these regimes was considered in details in the article [11]. Here we consider two other regimes which for our opinion are the most interesting ones from both theoretical and "consumer" points of view.

**The second-order-pole regime**   If the bandwidth of the Fabry-Perot resonator is small, $\gamma \ll \delta$, and the pumping energy is equal to[2]

$$\mathcal{E} \approx \frac{mL^2\delta^3}{8\omega_o}\left(1 + \frac{6\gamma^2}{\delta^2}\right), \tag{22}$$

then in the narrow vicinity of the frequency

---

[2]The exact expressions are too cumbersome so we present here only second-order Taylor expansions with respect to small parameter $\gamma/\delta$

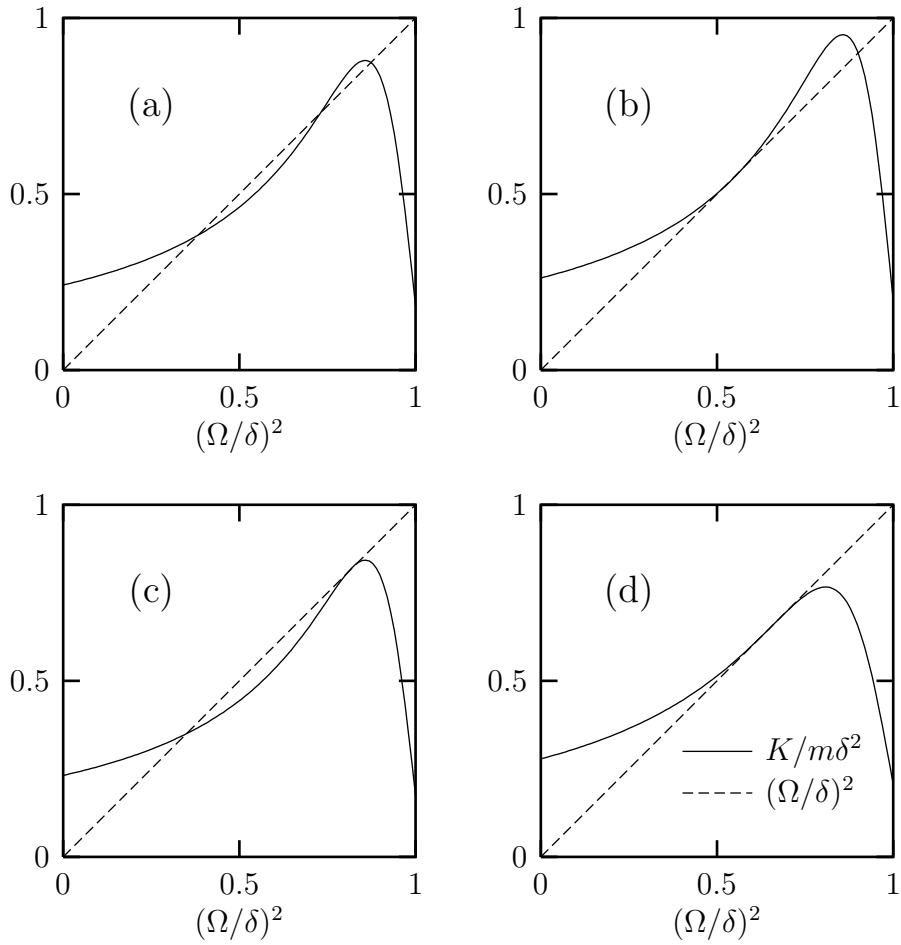

Figure 3: Different regimes of the frequency-dependent rigidity: (a) — three first-order poles; (b,c) — one second-order and one first-order poles; (d) — one third-order pole

$$\Omega_2 \approx \sqrt{\frac{\delta^2 + 11\gamma^2}{2}} \qquad (23)$$

the formula (17) can be presented as

$$S_{\text{total}}(\Omega) = \hbar m \delta^2 \left[ \frac{\gamma}{2\delta} + \frac{2\delta}{\gamma} \frac{(\Omega - \Omega_2)^4}{\Omega_2^4} \right]. \qquad (24)$$

The value of $\gamma/\delta$ can be adjusted to provide minimum of this spectral density at the edges of some given spectral band $\Omega_2 \pm \Delta\Omega/2$:

$$\frac{\gamma}{\delta} = \frac{1}{2}\left(\frac{\Delta\Omega}{\Omega_2}\right)^2. \qquad (25)$$

In this case there will be

$$S_{\text{total}}(\Omega_2 \pm \Delta\Omega/2) = \hbar m (\Delta\Omega)^2. \qquad (26)$$

This is the spectral equivalent of the formula (12).

In Fig. 4 spectral density of the total noise (17) is presented for several values of $\gamma$ and for the pumping energy (22) corresponding to the second-order pole (dashed line is the SQL level).

It is useful to compare sensitivity of this regime with the sensitivity provided by the usual probe oscillator with frequency independent rigidity and with eigen frequency $\Omega_m$. In the latter case (see article [10])

$$\xi^2 \equiv \frac{S_{\text{total}}(\Omega_m \pm \Delta\Omega/2)}{S_{\text{SQL}}(\Omega_m)} = \frac{\Delta\Omega}{\Omega_m}, \qquad (27)$$

where $S_{\text{SQL}}(\Omega) = \hbar m \Omega^2$ is the spectral density corresponding to the SQL for the free test mass. In the case of second-order pole regime will be

$$\xi^2 \equiv \frac{S_{\text{total}}(\Omega_2 \pm \Delta\Omega/2)}{S_{\text{SQL}}(\Omega_2)} = \left(\frac{\Delta\Omega}{\Omega_2}\right)^2 = \frac{2\gamma}{\delta}. \qquad (28)$$

**The third-order-pole regime** Parameters of the scheme can be tuned also to create the third order pole of the effective susceptibility (20) by setting

$$\mathcal{E} = \frac{9\sqrt{177} - 113}{49} \times \frac{mL^2\delta^3}{\omega_o} \approx \frac{0.14 mL^2\delta^3}{\omega_o} \qquad (29)$$

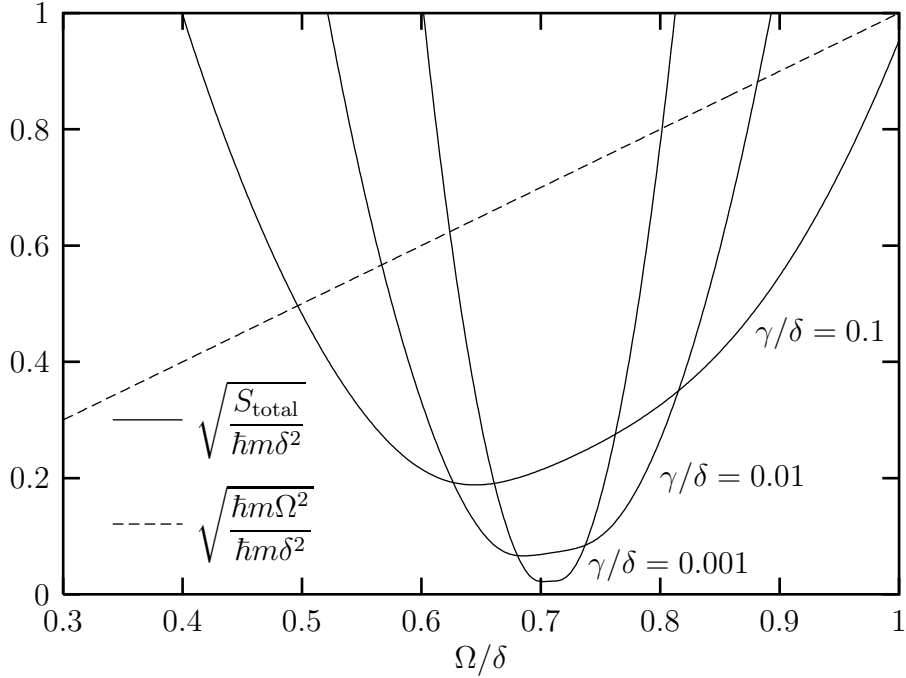

Figure 4: Sensitivity for the second-order-pole regime

and

$$\frac{\gamma}{\delta} = \frac{\sqrt{280 - 21\sqrt{177}}}{7} \approx 0.11 \,. \tag{30}$$

The pole frequency is equal to

$$\Omega_3 = \sqrt{\frac{22 - 2\sqrt{177}}{7}}\,\delta \approx 0.81\delta \,. \tag{31}$$

In Fig. 5 spectral density of the total noise (17) is presented for the case of the third-order-pole regime.

It is necessary to note that this third-order-pole regime is "overpumped": the second term in the formula (17), that is proportional to the measurement noise, is several orders of magnitude smaller than the first one (back-action noise) in the frequency area of interest. It is evident from Fig. 6, where these two terms are plotted separately. Usually, in such a situation the reduction of the total noise is possible by increasing the measurement noise and proportional decreasing the back-action noise due to using, for example, smaller value of the pumping energy. Unfortunately, within the framework of our simple scheme it is impossible because the same optical pumping is used both for measurement and for creating the rigidity. There is no additional "degree of freedom" here: values of all parameters are fixed by the equations (29,30). It is probable, however, that more sophisticated topologies based on use of separate optical modes for measurement and for creating rigidity, or/and which eliminate back-action noise by using variational measurement [5, 6, 15], will allow to create "well-balanced" third-order-pole regime with very low total noise.

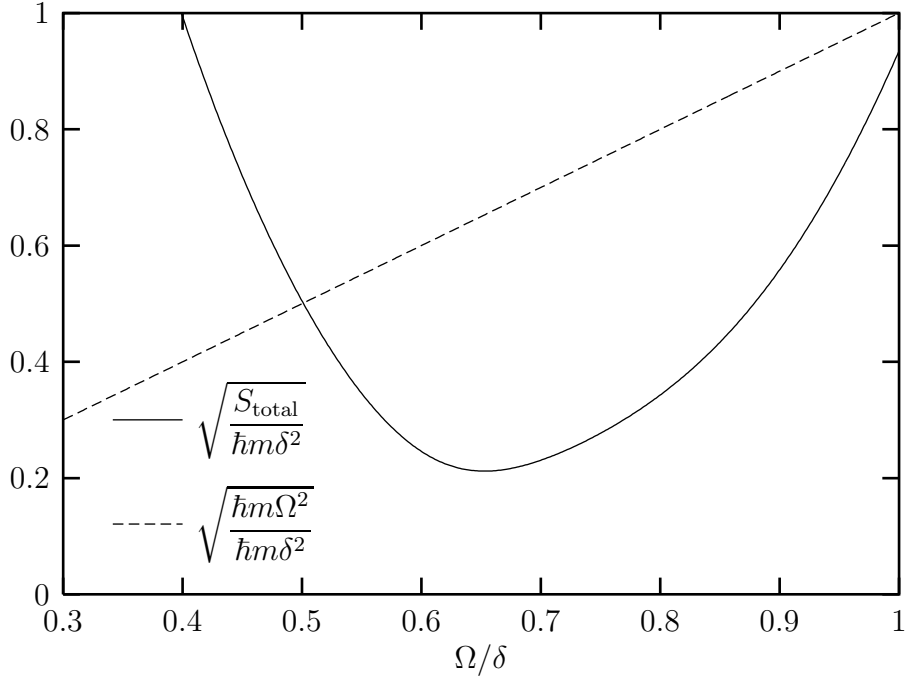

Figure 5: Sensitivity for the third-order-pole regime

## 4 Conclusion

It is evidently impossible to consider thoroughly in one short article all the possible variants of the use of the frequency-dependent rigidity which exists in large-scale optical resonators. Such a consideration has to be based on *a priori* information about the signal spectrum provided, for example, by astrophysical predictions in the similar as it had been done in the article [11]. It is evident, however, that:

- The second-order-pole regime allows to "dive" deep below the SQL in the narrow spectral band $\Delta\Omega$ which is, however, much wider than if the usual frequency independent rigidity is used [compare formulae (27) and (28)]. The recent achievements in fabrication of high-reflectivity mirrors [16] allows to expect that it will be possible to obtain relaxation time of the large-scale interferometers $\gamma^{-1} \gtrsim 1\,\text{s}$ and thus to reach the sensitivity at the level of $\xi^2 \lesssim 10^{-3}$, if $\Omega \sim \delta \sim 10^3\,\text{s}^{-1}$. It is important that the pumping energy in this regime does not depend on the sensitivity and remains approximately equal to the energy (15) which is necessary to achieve the SQL in the traditional scheme of the interferometric position meter.

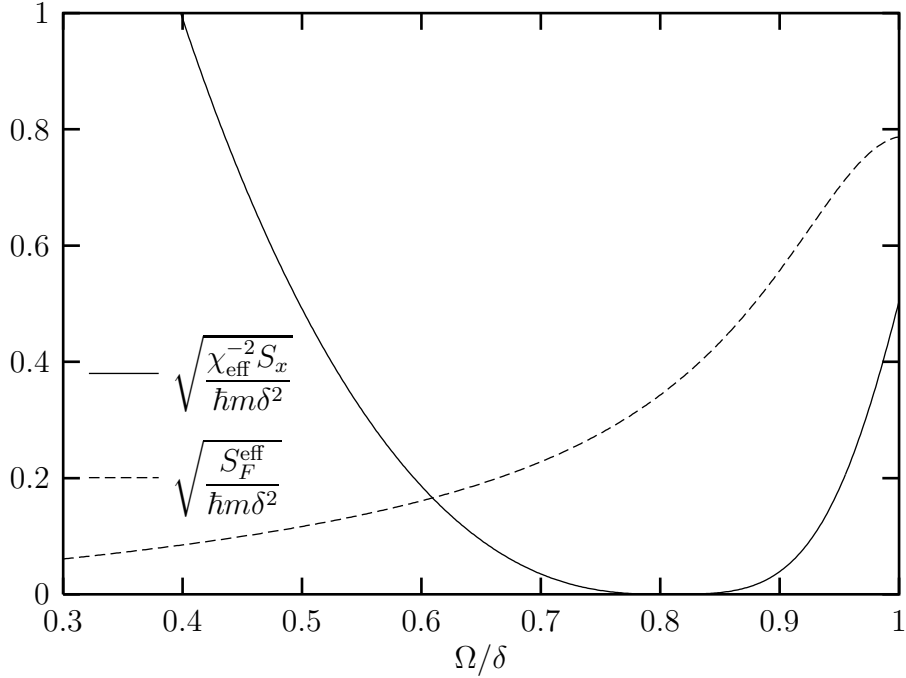

Figure 6: Measurement noise and back-action noise for the third-order-pole regime

- The third-order-pole regime provides sensitivity a few times better than the Standard Quantum Limit in relatively wide spectral band and at extremely low level of the measurement noise in this band. This regime looks as a good candidate for use in advanced topologies of the gravitational-wave antennae.

# Acknowledgments

Author thanks V.B.Braginsky and S.P.Vyatchanin for very useful discussions. This paper was supported in part by the California Institute of Technology, US National Science Foundation, by the Russian Foundation for Basic Research, and by the Russian Ministry of Industry and Science.